\documentclass{ws-ijmpe}

\newcommand{\mpi}{m_\pi}
\newcommand{\fpi}{f_\pi}
\newcommand{\mev}{\,{\rm MeV}}
\newcommand{\gev}{\,{\rm GeV}}

\newcommand{\eq}[1]{Eq.~(\ref{#1})}

\newcommand{\mpimax}{m_{\pi,\mathrm{max}}}

\begin{document}

\markboth{Young, Hall and Leinweber}{Towards selecting a finite-range
  regularization scale}

\catchline{}{}{}{}{}

\title{TOWARDS SELECTING A FINITE-RANGE REGULARIZATION SCALE}

\author{ROSS D.~YOUNG}

\address{Physics Division, Argonne National Laboratory, Argonne,
  Illinois 60439, USA.}

\author{JONATHAN M.~M.~HALL, DEREK B.~LEINWEBER}

\address{Special Research Centre for the Subatomic Structure of
  Matter, School of Chemistry and Physics, University of Adelaide,
  Adelaide, SA 5005, Australia.}

\maketitle

\begin{history}
\received{(received date)}
\revised{(revised date)}
\end{history}

\begin{abstract}
  Extensive studies have demonstrated that finite-range regularization
  (FRR) offers significantly improved chiral extrapolations for
  lattice QCD. These studies have typically relied on selecting the
  finite-regularization scale based upon phenomenological input.  Here
  we report on a preliminary investigation of a procedure to determine
  a preferred range of FRR scale based on nonperturbative lattice
  results --- without any phenomenological prejudice.
\end{abstract}

\section{Background}

Modern lattice QCD results are beyond the power-counting regime
(PCR)\cite{Beane:2004ks,Leinweber:2005xz} of the chiral expansion. Any
results from implementation of the chiral effective field theory (EFT)
in this domain are therefore model dependent. From the perspective of
lattice practitioners, the goal is focussed on {\em ab initio} studies
of nonperturbative QCD --- without phenomenological input. Thereby,
even when working outside the PCR, it can be attractive to choose
dimensional regularisation (or a similar minimal subtraction-{\em style}
scheme) to avoid phenomenological bias. As results are dependent upon
this choice of how the EFT is applied, then this application is
implicitly a model. Nevertheless, the {\em ab initio} demand is
maintained by not introducing additional information from {\em prior
  knowledge}.

Finite-range regularisation (FRR)\cite{Young:2002ib} has been
demonstrated to offer stable and robust chiral extrapolation of
lattice simulation results performed at moderate quark masses.  The
FRR procedure necessarily involves introducing a preference for a
regularisation scale that is ``low'' in favour of one that is infinite
--- whereby this assumption of prior knowledge means one can no longer
assert {\em ab initio}.

Here we highlight an initial investigative study in search of a
technique which will enable the results of lattice simulations to
select a preferred regularisation scale. By removing the need to
artificially select the regularisation scale by hand, one can utilise
the benefits of FRR and maintain {\em ab initio} status.

\section{Convergence and the power-counting regime}
The the success of effective field theories, such as chiral
pertubation theory, lies in the presence of a small expansion
parameter and the reliance of a series expansion of natural-sized
coefficients. A simple example function is the geometric series
\begin{equation}
\frac{1}{1+x} = 1-x+x^2-x^3+\ldots
\label{eq:geo}
\end{equation}
The radius of convergence of this series expansion about $x=0$ is
$|x|<1$. Given a truncated expansion, the deviation from the exact
result, at a given value of $x$, can be estimated by considering the
size of the first neglected term in the series --- without necessarily
having knowledge of the full result.

This toy example can provide insight into the natural
convergence radius of the quark-mass expansion of the nucleon
mass. This expansion can be written as
\begin{equation}
M_N=M_N^0+c_2\mpi^2+\chi_\pi\mpi^3+\ldots\,,
\label{eq:MNLNA}
\end{equation}
where $\mpi\propto \sqrt{m_q}$. The $\mpi^3$ term is nonanalytic in
the quark mass (ie. $\sim m_q^{3/2}$), and it's coefficient is know
model-independently,
\begin{equation}
\chi_\pi = -\frac{3\,g_A^2}{32\pi\fpi^2}\,.
\label{eq:chipi}
\end{equation}
Strictly, $g_A$ and $\fpi$ are to be evaluated in the chiral
limit. Assuming the physical values do not differ too significantly,
this suggests a numerical value of $\chi_\pi\simeq-5.6\gev^{-2}$.  The
term in $\mpi^2$ is also known phenomenologically through the
pion-nucleon sigma term, whereby\cite{Gasser:1990ce}
\begin{equation}
\sigma_N=\mpi^2\frac{d\,M_N}{d\,\mpi^2}\simeq 45\mev\,.
\end{equation}
Using this and the physical nucleon mass, together with the leading
nonanalytic term gives, $c_2\simeq 3.5\gev^{-1}$ and $M_N^0\simeq
0.89\gev$.

Using these numerical estimates in \eq{eq:MNLNA} one finds that the
second and third term reach 100\% of the leading term at pion masses
of $0.51\gev$ and $0.54\gev$, respectively.  This suggests that at a
pion mass of the order $\mpi\sim 0.5\gev$ one is at the radius of
convergence of this series --- where an infinite number of terms are
required to reproduce the exact result. Of practical importance is how
precisely can the curve be reproduced with a finite number of terms.

The first few terms of the expansion \eq{eq:MNLNA} looks very much like
the geometric series, setting $m_R=0.54\gev$,

\begin{equation}
M_N=\left(1+1.1\left(\frac{\mpi}{m_R}\right)^2
-1.0\left(\frac{\mpi}{m_R}\right)^3+\ldots\right)\left[0.89\gev\right]\,.
\label{eq:MNseries}
\end{equation}
The suggests that using \eq{eq:geo} with $x=\mpi/m_R$ (ignoring the
term linear in $x$) is a good way to estimate the precision of a given
truncation.  Demanding that the series expansion in \eq{eq:geo} up to
$x^3$ is accurate to the 1\% level, this limits one to $x<0.34$. Thereby
1\% precision is enabled up to $\mpi/m_R<0.34$ or $\mpi<0.18\gev$.
Continuing the natural-size argument means that at order $\mpi^4$
demanding this level of precision restricts one to
$\mpi<0.23\gev$. Expansions to higher order in the nucleon mass have
also recently been revisited\cite{McGovern:2006fm,Schindler:2006ha},
but even going to $\mpi^5$ or $\mpi^6$ still only provides a valid
expansion up to $\mpi<0.27$ and $0.30\gev$, respectively. Further,
going to increasing order also comes at the cost of fitting more
parameters and thereby requiring even more simulation results below
the quoted thresholds.

\section{Numerical approach to FRR scale determination}
With the aim of performing precision {\em ab initio} studies, it is
evident that lattice results from beyond the (1\%) PCR must be used.
The first option would be to just
to use the truncated forms of a DR expansion. In using lattice results
up to $\mpi\sim 0.5\gev$ to constrain an extrapolation based upon a
fourth-order expansion would lead to a systematic uncertainty of 35\%
(based on the geometric series discussed above). A second option is to
introduce a finite-range regularisation scale with the aim of both
incorporating more data, to improve statistical precision, and
minimising the systematic uncertainty.

The physical interpretation for the success of FRR lies in the
suppression of rapidly-varying chiral logs at moderate quark
masses. Mathematically, the success is based on using a separation
of scales to introduce an effective resummation of higher-order terms
in the chiral series, while maintaining the model-independence of the
expansion to the order one is working.

To continue with the nucleon mass example, the chiral expansion,
\eq{eq:MNLNA}, can be rewritten in an unrenormalised form
\begin{equation}
M_N=a_0+a_2\mpi^2+a_4\mpi^4+\Sigma(\mpi,\Lambda)\,,
\label{eq:MNbare}
\end{equation}
where $\Sigma$ denotes the meson-loop corrections, which contains the
chiral nonalyticities, and $a_i$ are regularisation-scale dependent
expansion coefficients. In adjusting the regularisation scale
$\Lambda$, one can shift strength bewteen the loop contributions and
the residual series composed of the $a_i$'s. FRR works well because
with an appropriately chosen $\Lambda$, the residual series 
typically shows much better convergence than the renormalised form.
We wish to exploit this feature in using lattice results to choose an
``optimal'' range for $\Lambda$.

Further, any criteria to determine $\Lambda$ should produce certain
desired limits. Firstly, if one really is using data in the PCR, then
there should be no pereference for a finite $\Lambda$, and
$\Lambda\to\infty$ should display equally good covergence
properties. In fact, the data should always reject a $\Lambda$ which
is too low. In the PCR, then letting $\Lambda\to 0$ would give a curve
inconsistent with the nonanalyticities of the data. Further, if the
expansion is extended to quark masses beyond the range accessible
within FRR, then a solution consistent with $\Lambda\to 0$ is a clear
indication that the data is inconsistent with the chiral behaviour.
In this domain, it may also be evident that no amount of tuning
$\Lambda$ can stablize the convergence.

The technique to constrain $\Lambda$ we investigate here, is to
numerically adjust $\Lambda$ such as to ensure the {\em residual}
series is satisfactorily converging over the range upon which the
lattice results are fit. In the case above, \eq{eq:MNbare}, a
preferred value of $\Lambda$ is one which gives numerical evidence of
convergence of the series
\begin{equation}
\left\{a_0\,, a_2 \mpimax^2\,, a_4 \mpimax^4\,,\ldots\,  \right\}\,,
\end{equation}
where $\mpimax$ denotes the largest pion mass at which the expansion
in used in a fit. A
potential convergence criteria is to ensure that
\begin{equation}
\mathcal{R}\equiv \left| \frac{a_4\mpimax^4}{a_2\mpimax^2} \right|
\label{eq:R}
\end{equation}
remains small.  Here, \eq{eq:MNbare} is assumed to be working to just
leading nonanalytic order, ie. fully renomalised to ${\mathcal
  O}(\mpi^3)$. The $\mpi^4$ term is introduced in the fit to reduce
the sensitivity to the ultra-violet cutoff\cite{Young:2002ib}. It is
desired that this term only contribute of the order 1\% of the total
result --- corresponding to roughly 10\% of the $\mpi^2$ term,
suggesting an initial test scale of $\mathcal{R}<0.1$.


\section{Rho-meson in quenched-QCD}
Quenched QCD offers a useful framework to test methods and techniques
where the cost of dynamical simulations remain too computationally
expensive. Further, in the case of the rho-meson, neglecting the
dynamical sea quarks removes the decay channel of the rho. Thereby,
one does not have to deal with complications arising through
extrapolating through the 2-$\pi$ decay threshold\cite{Allton:2005fb}.

On the assumption that disconnected contributions in the $\omega$
propagator are negligible, the only leading one-loop diagrams
contributing to the $\rho$ mass are those associated with the
flavor-singlet $\eta'$. The $\eta'$ only appears in the low-energy EFT
as a result of quenching, whereby it remains degenerate
with the Goldstone pion.

Our
expansion for the rho-meson mass in quenched QCD is summarized as
\begin{equation}
m_\rho = \sqrt{\hat{m}_\rho^2+\Sigma}
\label{eq:rho}
\end{equation}
where the residual, unrenormalised expansion is defined by
\begin{equation}
\hat{m}_\rho = a_0 + a_2 \mpi^2 + a_4 \mpi^4
\end{equation}
and the chiral loop corrections are denoted by $\Sigma$.
%
Further details will appear in a forth-coming
manuscript\cite{Hall:2008}.

\section{Preliminary investigation}
Here we wish to test the hypothesis to determine $\Lambda$ by
constraining the quantity $\mathcal{R}$, \eq{eq:R}. To do so, we
construct some test, or pseudo, data in order to perform hypothetical
fits. Firstly, we fit \eq{eq:rho} to {\em new, unpublished,} quenched,
overlap results of the Kentucky Group\cite{KY:2008} using a dipole
regulator of scale $\Lambda=0.8\gev$.  This input simply enables a
scale to be set for the test data. A more detailed study would
investigate alternative constraints at this initial step.

With this fit we construst a dense set of pseudo-data between the
physical pion mass and some upper value $\mpimax$. We then refit this
pseudo-data with alternative regularization scales and assess the
convergence criteria. The convergence test is plotted for varying
$\Lambda$ for a test curve which using an upper $\mpimax=0.3\gev$.
\begin{figure}[th]
\centerline{\psfig{file=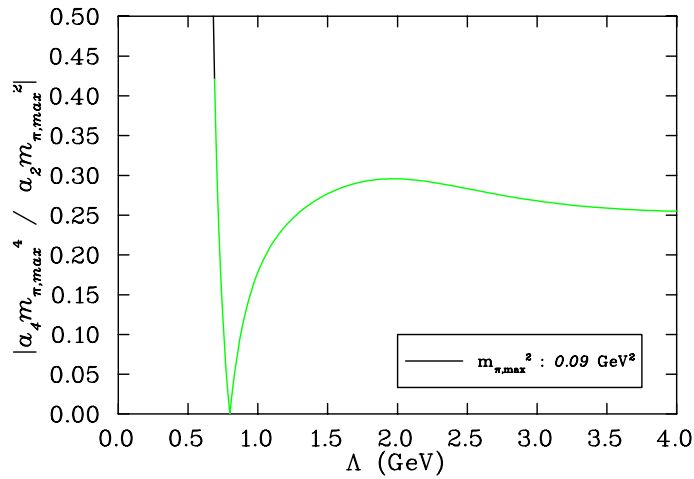,angle=90,width=9cm}}
\vspace*{8pt}
\caption{Convergence criteria, $\mathcal{R}$, plotted against regularisation scale.}
\label{fig:R}
\end{figure}

The first indication of support for the procedure is the failure to
produce a converging series if the regularisation scale is chosen as
too small. This reflects the fact that the analytic expansion cannot
describe the analyticities of the data. The existence of a lower bound
as determined by this criteria is then a signature that the chiral
nonanalyticities are present in the data.

There is a window of $\Lambda$ values where the series is
satisfactorily converging to meet the demands of 1\% precision. This
happens to be in the vicinity of the originally input scale of
$\Lambda=0.8\gev$ and success of the fit should come as no
surprise. From our initial tests using pseudo-data constrained using
$\Lambda=2.0\gev$, this convergence criteria reflects this and chooses
an optimal regulator in this vicinity. Thereby, with good enough data,
the intrinsic preferred scale can reveal itself through this criteria.

The curve indicates that should a large regularisation scale be
attempted to fit our pseudo-data, the residual series is not so well
converging. Even here, with an upper pion mass of $300\mev$, truncation of the 
residual series suggests a $(0.25)^2$ or 6\% uncertainty.

Another desirable feature of this procedure is that as $\mpimax$ is
reduced, the sensitivity to $\Lambda$ must disappear through the
additional factor of $\mpimax^2$ in $\mathcal{R}$. {\em Thereby, within the
PCR the lower bound will remain, yet no significant upper bound will
persist.}

On face value, this looks like a reasonable procedure to allow lattice
data itself to make an {\em ab initio} determination of an optimal FRR
scale. A weakness is apparent in considering the renormalisation of
the residual series. For this argument, considering just the
renormalisation of the single-hairpin graph. The Taylor expansion of
this diagram, up to normalisation, can be expressed as\cite{Young:2002ib}
\begin{equation}
b_0 + b_2 \mpi^2 + \mpi^3 + b_4 \mpi^4 + \ldots
\end{equation}
The leading term $b_0$ behaves as $\Lambda^3$, while $b_2\sim\Lambda$
and $b_4\sim\Lambda^{-1}$. Since for any fit the renormalised
parameters are essentially stabilised by the data, this indicates that
for very large $\Lambda$ the residual series coefficients will behave
as $a_{\{0,2,4\}}\sim \Lambda^{\{3,1,-1\}}$. Thereby, $a_2$ will
diverge and $a_4$ saturate to a constant, and consequently our
described convergence criteria $\mathcal{R}$ will approach zero,
regardless of the data.

While a promising approach, it appears that the criteria described by
\eq{eq:R} does not lead to a conprehensive test of convergence. A
potential modification could be to normalise the $a_4 \mpimax^4$ term
to a renormalised quantity, such as the rho mass. We anticipate
further studies in this direction will lead to a reliable scale
determination procedure that will facilitate {\em ab initio} studies
with FRR for all $\Lambda$.

\section*{Acknowledgements}
We thank C.~D.~Roberts for discussions and K.~F.~Liu and N.~Mathur of
the Kentucky Group for sharing their preliminary lattice simulation
results used in this investigation.  This work was supported by the
Department of Energy, contract no.~DE-AC02-06CH11357 and the
Australian Research Council.


\begin{thebibliography}{20}

\bibitem{Beane:2004ks}
  S.~R.~Beane,
  Nucl.\ Phys.\  B {\bf 695}, 192 (2004)
  [arXiv:hep-lat/0403030].

\bibitem{Leinweber:2005xz}
  D.~B.~Leinweber {\it at al.},
  Nucl.\ Phys.\  A {\bf 755}, 59 (2005)
  [arXiv:hep-lat/0501028].

\bibitem{Young:2002ib}
  R.~D.~Young {\it et al.},
  Prog.\ Part.\ Nucl.\ Phys.\  {\bf 50}, 399 (2003)
  [arXiv:hep-lat/0212031].

\bibitem{Leinweber:2003dg}
  D.~B.~Leinweber {\it at al.},
  Phys.\ Rev.\ Lett.\  {\bf 92} (2004) 242002
  [arXiv:hep-lat/0302020].

\bibitem{Allton:2005fb}
  C.~R.~Allton {\it at al.},
  Phys.\ Lett.\  B {\bf 628} (2005) 125
  [arXiv:hep-lat/0504022].

\bibitem{Li:1971vr}
  L.~F.~Li and H.~Pagels,
  Phys.\ Rev.\ Lett.\  {\bf 26}, 1204 (1971).

\bibitem{Gasser:1990ce}
  J.~Gasser, H.~Leutwyler and M.~E.~Sainio,
  Phys.\ Lett.\ B {\bf 253}, 252 (1991).

\bibitem{McGovern:2006fm}
  J.~A.~McGovern and M.~C.~Birse,
  Phys.\ Rev.\  D {\bf 74}, 097501 (2006)
  [arXiv:hep-lat/0608002].

\bibitem{Schindler:2006ha}
  M.~R.~Schindler, D.~Djukanovic, J.~Gegelia and S.~Scherer,
  Phys.\ Lett.\  B {\bf 649}, 390 (2007)
  [arXiv:hep-ph/0612164].

\bibitem{Hall:2008}
  J.~M.~M.~Hall, D.~B.~Leinweber and R.~D.~Young, {\it in preparation}.

\bibitem{KY:2008}
  K.~F.~Liu and N.~Mathur, {\em preliminary results} of the Kentucky Group.

\end{thebibliography}
\end{document}